\documentstyle[preprint,aps]{revtex}

\begin{document}
\preprint{\begin{tabular}{c}
\hbox to\textwidth{June 1994 \hfill SNUTP 94-55}\\[-10pt]
\hbox to\textwidth{hep-ph/9406296 \hfill TUM-HEP-203/94}\\[36pt]
\end{tabular}}

\title{\bf Symmetry Principles
toward Solutions of the $\mu$ Problem}

\author{Jihn E. Kim$^{(a)}$ and Hans Peter Nilles$^{(b,c)}$}

\address{\sl
$^{(a)}$Department of Physics and
Center for Theoretical Physics\\[-8pt]
Seoul National University\\[-8pt]
Seoul 151-742, Korea\\
$^{(b)}$Physik Department, Technische Universit$\ddot {\sl a}$t
M$\ddot {\sl u}$nchen\\[-8pt]
D-85747 Garching, Germany\\
$^{(c)}$Max-Planck-Institut f$\ddot {\sl u}$r Physik,
Werner-Heisenberg-Institut\\[-8pt]
D-80805 M$\ddot {\sl u}$nchen, Germany}

\maketitle

\def\be{\begin{equation}}
\def\ee{\end{equation}}

\begin{abstract}
We stress that a natural solution of the $\mu$ problem requires two
ingredients: a symmetry that would enforce $\mu = 0$ as well as the
occurence of a small breaking parameter that generates a nonzero $\mu$.
It is  suggested that both the Peccei-Quinn symmetry and
the spontaneously  broken $R$ symmetry may be the sources of the
needed $\mu$ term in the minimal supersymmetric standard model
provided that they are  spontaneously broken at
a scale $10^{10}-10^{12}$ GeV.  To
solve the strong CP problem with a hidden
sector confining group, both of these symmetries are needed
in superstring models with an anomalous $U(1)_A$.
\end{abstract}
\pacs{}

%\narrowtext

\noindent {\bf 1. Introduction}

It is  often assumed
that one can
 understand many aspects of the low energy electroweak phenomena
from supergravity interactions at high energies governed by an
underlying superstring theory \cite{gsw}.  Even though
there does not exists a {\it standard} superstring model,
it is a general expectation that one of numerous string vacua
might  coincide with the minimal supersymmetric standard model (MSSM)
in supergravity \cite{nhk}. Nevertheless, the attractive MSSM
has a few outstanding theoretical
parameter problems: the cosmological constant problem \cite{ckn1},
the strong CP problem \cite{rev}, and the $\mu$ problem \cite{kn}.
{}From the top--down approach, these may be solved by the $theory$.
But from the bottom--up approach, these need explanations
and constitute very interesting
and challenging problems.  In this paper, we address two
of these problems: the strong CP problem and the $\mu$
problem.

The strong CP problem seems to have an inherent solution in
string models due to the existence of the model-independent
axion \cite{mia}.  However, the model-independent axion
is known to have the axion decay constant problem if it is
designed to solve the strong CP problem \cite{constant}.
Apart from this cosmological  problem, the need for a hidden sector
confining group to break supersymmetry at the intermediate
scale requires another axion to settle the QCD vacuum
angle at zero.  However, the string models are known to
have no continuous global symmetry.  Therefore, there does not exist any room
for the additional axion at the string level.  Thus the
prospect for a solution of the strong CP problem
through an invisible axion looks doomed.

On the other hand, the MSSM admits
\be
W_\mu = \mu H_1H_2
\ee
in the superpotential,
where $H_1$ and $H_2$ are two Higgs doublets with $Y=-1/2$
and $Y=1/2$ and
$\mu$ is a free parameter whose magnitude is in principle
different from the magnitude of $m_{3/2}\sim M_{SUSY}$
characterizing the soft SUSY breaking terms in supergravity.
Electroweak phenomenology suggests that $\mu$ is nonzero
and falls in the 100 GeV region similar to the magnitude
of soft parameters.  In addition,
axion phenomenology suggests that $\mu$
cannot be zero, or there results an unwanted 0.1 MeV axion.
Theoretically, however, one expects that $\mu$ is of order
the Planck scale unless it is forbidden by some symmetry.
At present, there exist several suggestions toward a solution
of the  $\mu$ problem
\cite{{kn},{gm},{munoz}}.

In this paper, we will try to understand these two
outstanding problems from symmetry principles, which
will work as a guideline for future construction of
superstring models.

\noindent {\bf 2. The Peccei--Quinn and $R$ Symmetries}

{}From the early days of the $\mu$ problem \cite{kn},
it was suggested that
a symmetry, presumably a Peccei-Quinn symmetry with an invisible
axion \cite{axion}, may be needed toward a solution of the
problem.  The reason is simple.  Consider the following Yukawa
super-potential
\be
W_Y = d^{cT}_L F_d q_LH_1+ u^{cT}_L F_u q_LH_2
\ee
where $F_{d,u}$ is the $3\times 3$ Yukawa coupling matrix, $q_L$
is a column vector of three SU(2) doublet superfields,
$d^c_L$ and $u^c_L$ are column vectors of up- and down-type
anti-quark superfields. A Peccei--Quinn symmetry\footnote{For
the heavy quark axion, $H_{1,2}\rightarrow H_{1,2}$, and
hence it cannot be used for the small $\mu$ term.}
\begin{eqnarray}
H_1\longrightarrow e^{i\alpha}H_1,\ \
H_2\longrightarrow e^{i\alpha} H_2&&,\ \
\{q_L, d^c_L, u^c_L\}\longrightarrow e^{-{i\over 2}
\alpha}\{q_L, d^c_L, u^c_L\}\nonumber \\
W&&\longrightarrow W.
\end{eqnarray}
forbids $W_\mu$.
One can introduce the Peccei--Quinn symmetry to understand the
smallness of $\mu$, and break it at $10^{10}-10^{12}$ GeV,
generating an electroweak scale $\mu$ through nonrenormalizable
interactions generated by gravity \cite{kn}.
The Peccei--Quinn symmetry seems to be a necessity to fix
the scale of $\mu$ at $M_{SUSY}$.

In this paper, we also suggest that $R$ symmetry can be used as
another symmetry in addition to the Peccei--Quinn
symmetry.\footnote{The use of $R$ symmetry
for the $\mu$ term has been suggested before \cite{{ckn},{dm}}.}
With $\mu=0$, we assign an $R$ symmetry,
\begin{eqnarray}
\{H_1, H_2\}\longrightarrow &
\{H_1, H_2\},\nonumber \\
\{q_L, d^c_L, u^c_L\}\longrightarrow &e^{i\beta}
\{q_L, d^c_L, u^c_L\},\\
W\longrightarrow &e^{2i\beta}W\nonumber
\end{eqnarray}
so that ordinary quarks and Higgs doublets carry vanishing $R$
charge.

\noindent {\bf 3. $R$ Violation and the $\mu$ Term}

As a simple example, consider the Polonyi super-potential
\cite{polonyi},
\be
W_{Polonyi}=m^2(z+\tilde m)
\ee
where $\tilde m$ is of order of the Planck scale
($M=M_{Pl}/\sqrt{8\pi}$) and $m$ is of the order of intermediate
SUSY breaking scale ($M_{I}$). Supersymmetry is broken at the scale
$m(\sim M_{I})$, $F_z=m^2+(m^2/M^2)z^*(z+\tilde m)$ with
$z=(\sqrt{3}-1)M$ and
$\tilde m=(2-\sqrt{3})M$.  With $\tilde m=0$, $W=W_Y+W_{Polonyi}$
has the $R$ symmetry with $z$ transforming under $R$ as
\be
z\longrightarrow e^{2i\beta}z.
\ee
The field $z$ gets mass through  $m$ of order
$ O(m^2/M)=O(m_{3/2})$.  From the supergravity Lagrangian, we note
\cite{ferrara} that the $z$ field interacts with the other observable
sector fields through gravitation only, and hence it has the severe
cosmological problem \cite{kolb}.  The hypothetical $R$ symmetry is
broken by nonzero $\tilde m$. To see the effect of $R$ symmetry
violation, we can assign $R$--charge 2 to $m^2\tilde m$.  Thus, we
expect that this model generates through gravitational
interaction a super-potential $(m^{2}\tilde m)H_1H_2/M^2$,
producing $\mu\sim m_{3/2}$.
Any supergravity model, with super-potential $W={\rm cubic\ terms}\ +\
$constant, realizes this kind of $R$ breaking and accordingly
generates a nonzero $\mu$ term.

Let us briefly discuss the models discussed by Guidice and Masiero
\cite{gm} and by
Casas and Munoz \cite{munoz} from this $R$ symmetry argument.
Guidice and Masiero suggest a K$\ddot {\rm a}$hler potential
\be
K= (H_1H_2+h.c.)+\cdots
\ee
where $\cdots$ denotes $\phi^*\phi$, etc.  In supergravity models,
the Guidice and Masiero solution gives an $H_1H_2$ term in the
scalar potential, not to the superpotential.  However, its effect
can be absorbed in the superpotential through the transformation,
\be
K(\phi,\bar\phi)\longrightarrow K(\phi,\bar\phi)-F(\phi)-\bar F(\bar
\phi),\ \ \ W(\phi)\longrightarrow e^{F(\phi)}W(\phi).
\ee
Therefore, the solutions discussed in Refs. \cite{{gm},{munoz}} can
be studied in the framework of Ref. \cite{munoz}.  Casas and Munoz
assume that the superpotential $W_0$ does not contain the $\mu$
term.  But the gravitational interaction can generate a term
$W_0H_1H_2/M^2$.  They assume $\langle W_0\rangle\sim
M^2m_{3/2}$ to generate $\mu\sim M_W$
and relate it to the scale of supersymmetry breakdown.\footnote{In this
model there is no symmetry that forbids  the $\mu$-term and dangerous
renormalizable operators like,
 for example, $Z_iH_1H_2$ in $W_0$ with nonvanishing
vacuum expectation values of gauge singlets $Z_i$. Thus, a full
solution of the $\mu$ problem is not achieved.}  The effective
superpotential $W_0H_1H_2/M^2$ preserves the $R$ symmetry, and
$\langle W_0\rangle\ne 0$ breaks the $R$ symmetry and there results
a pseudo-Goldstone boson, which is hidden in the fields describing
$W_0$.  This solution may have the Polonyi problem too \cite{kolb}.

A more interesting case is provided in models with hidden sector
confining gauge groups \cite{nilles}.\footnote{The most popular
attempts of supersymmetry breaking
in superstring models rely on this idea \cite{din}.}
The $\mu$ term generated in this scenario has been discussed in
Ref. \cite{ckn}.  We will elaborate this mechanism
after presenting general classifications of $R$ breaking mechanisms.

In string derived supergravity models,
there exist only cubic terms in the superpotential.  One can then
define an $R$ symmetry.  This $R$ symmetry could be broken
\footnote{In such theories a
pseudo-Goldstone boson $\chi$ does appear.  Its mass can be estimated by
standard perturbative methods.} in various ways:
(a) vacuum expectation values of $R\ne 0$ scalar fields, (b) $F$ terms of
$R=0$ chiral fields, (c) a constant in the effective superpotential,
or (d) gaugino condensation.
\vskip 0.4cm

\noindent {\it Vacuum expectation values of $R\ne 0$ scalar fields}
-- Let the chiral fields carrying nonzero $R$ quantum numbers be
$B_i$ with $R=R_i$.  Then the effective superpotential of the form
\be
{1\over M^{n-1}}[\prod_{i=1}^n B_i]H_1H_2,\ \ \ \sum_{i=1}^n R_i=2
\ee
preserves the $R$ symmetry.  The $R$ symmetry is broken by the
vacuum expectation values of $B_1, B_2, \cdots ,$ and $B_n$, and
$\mu$ is generated
\be
\mu={\langle B_1B_2\cdots B_n\rangle\over M^{n-1}}.
\ee
For the argument of $R$ symmetry toward a naturally small $\mu$ to make
any sense, $\langle B_i\rangle\ll M$.
\vskip 0.4cm

\noindent {\it F terms of chiral fields} -- Let us suppose that
$A_i$ carry vanishing $R$ quantum numbers so that $A_{iF}$
carries $R=-2$.  Then $A_i^*H_1H_2$  needed for the $D$-term
does not carry $R$ quantum number and generates a $\mu$ term
\be
{1\over M}\int d^2\bar\theta d^2
\theta A_i^*H_1H_2\ =\ \int d^2\theta W_\mu
\ee
where $\mu=A^*_{iF}/M\sim m_{3/2}$.  The dilaton
superfields, moduli fields and chiral fields
corresponding to flat directions can have vanishing $R$ quantum
numbers, since they do  appear in the superpotential only
as multiplicative factors of terms involving the interaction
of the matter fields.  For $B_i$
fields carrying $R=\pm 1$, one needs two $B$'s,
e.g. $B_1(R=1)$ and $B_2(R=-1)$ to have a $D$-term,
\be
{1\over M^2}\int d^2\bar\theta d^2\theta B_1^*B_2^*H_1H_2.
\ee
In this case,
\be
\mu\sim (1/M^2)(B^*_{1F}\langle B^*_2\rangle
+\langle B^*_1\rangle B^*_{2F}).
\ee
Since $\langle B_i\rangle\ll M$ for $R$ symmetry to solve the
$\mu$ problem, the $\mu$ term generated by $B_i$'s are insignificant
compared to the one arising from the $A_i$ terms.
\vskip 0.4cm

\noindent {\it Constants in $W_{\rm eff}$} -- As discussed above the
constant in the superpotential breaks the $R$ symmetry.  In string
motivated supergravity models, this constant arises from the
vacuum expectation values of $A_iB_jB_k$ where $R(A_i)=0$ and
$R(B_j, B_k)=1$ or more generally $R(A_i)+R(B_j)+R(B_k)=2$.
For this to generate a gravitino mass scale
$\mu$, we require
\be
\langle A_i\rangle \sim M,\ \ \
\langle B_i\rangle \sim M_{I}.
\ee
Thus to obtain a reasonable $\mu$, the scalar components of
chiral fields participating in the cubic superpotential with
nonzero $R$ should have vacuum expectation values not exceeding
the intermediate scale.
\vskip 0.4cm

\noindent {\it Gaugino condensation} -- Gauginos are assigned with
$R=+1$ so that gauge bosons carry the vanishing $R$  quantum number.
Therefore, the gaugino condensation in the hidden sector breaks
supersymmetry and can generate a $\mu$ term through
\be
{1\over M^2}\langle \Psi^a\Psi^a\rangle H_1H_2
\ee
where the (chiral) gauge boson multiplet is $\Psi^a=i\lambda^a
+\cdots$.\footnote{One can obtain the same effect by
generalizing the gauge kinetic function \cite{choi}.}
The magnitude of $\mu$ is
\be
\mu\sim {\Lambda_h^3\over M^2}.
\ee
\vskip 0.4cm

\noindent {\bf 4. Solutions of the Strong CP and $\mu$
Problems in String Models}

This leads us to the discussion on the anomalous $U(1)$ symmetry
\cite{anom} in string models and generation of the $\mu$ term
through the Peccei--Quinn symmetry \cite{ckn}.  The anomalous
gauge $U(1)$ becomes a global $U(1)_P$ by removing the massive
anomalous gauge boson\cite{kim}.\footnote{The reason
that the global symmetry results is similar to that an $SU(2)$
global symmetry results if an $SU(2)$ gauge symmetry without
Yukawa couplings is broken
by the vacuum expectation value of a doublet Higgs field.}
It obtains mass by absorbing the model-independent axion.
Frequently, string models are referred to having no
global symmetry, which led some to assume an approximate
$U(1)_{PQ}$ in some string models \cite{ps}.  However, the class
of string models with an anomalous $U(1)_A$ have
one $U(1)_P$ global symmetry
below the string scale for a solution of the strong CP problem.
The $U(1)_P$ symmetry has the $P$--(gauge boson)--(gauge boson)
anomaly,\cite{{kim},{ckn}},
\be
\partial_\mu J^\mu_P={1\over 32\pi^2}\sum_{G_i}F_{\mu\nu}(G_i)
\tilde F^{\mu\nu}(G_i)
\ee
where $G_i$ is the gauge group.  With one nonabelian gauge
group \cite{kim}, $U(1)_P$ is the needed Peccei--Quinn symmetry
for the solution of the strong CP problem.  But the
hidden sector confining group needed for SUSY breaking
invalidates the strong CP solution
by the invisible axion since the axion in general gets a mass of
order (coupling)$\times \Lambda_h^2$ where $\Lambda_h$ is the
scale of the hidden sector confining gauge group.
In Ref. \cite{ckn}, it was pointed out that $U(1)_R$ can be
used to obtain the invisible axion.
We elaborate briefly how this idea can be made successful.

With $U(1)_P$ and $U(1)_R$ symmetries, one may consider a
potential of the form
\be
V\ =\ -\Lambda_{QCD}^4\cos (\theta_c+\nu_0\theta_h)-
\Lambda^4_h\cos (\theta_c+\nu_h \theta_h)+V_R(\theta_c,\theta_h)
\ee
where $\theta_c(=a_c/v_c)$ is the QCD vacuum angle and $\theta_h
(=a_h/v_h)$ is the hidden sector vacuum angle.  The Goldstone bosons
corresponding to these symmetries are $a_c$ and $a_h$ with
decay constants $v_c$ and $v_h$, respectively.
We assume that $U(1)_R$
has the $R-SU(3)_c-SU(3)_c$ and $R-G_h-G_h$ anomalies which
are parametrized by $\nu_0$ and $\nu_h$, respectively. The
$U(1)_P$ has both $U(1)_P-SU(3)_c-SU(3)_c$ and $U(1)_P-
G_h-G_h$ anomalies.  Without $V_R$, both $\theta_h$ and $\theta_c$
are settled to zero.  One can expect the appearance of
nontrivial $V_R$, and there exists a possibility that $\theta_c$
is not settled to zero.  However, to study the extra potential
$V_R$, one must consider the original $U(1)_P\times U(1)_R$
symmetry.  Let us consider the following chiral and gauge
fields as a simple example,
\begin{equation}
\begin{array}{cccccc}
&A& B& \Psi^a& H_1& H_2\\
P\ \  &2& -1& 0& 1& 1\\
R\ \  &0& 1& 1& 0& 0
\end{array}
\end{equation}
where $P$ and $R$ are the charges of $U(1)_P$ and $U(1)_R$.
The potential $V_R$ arises from effective super-potential
\be
\int d^2\theta W_{\rm eff}\ =\ \int
d^2\theta\{ W_1+\int d^2\bar\theta
g_1\bar g_2\}
\ee
where $W_1$ and $g_1$ are the composite superfield operator
constructed from (left-handed) chiral fields only and $\bar g_2$
is constructed from (right-handed) anti-chiral fields only.  From
the $U(1)_P\times U(1)_R$ symmetry, we argue that $W_1$ carries
vanishing $P$ and 2 units of $R$ while $g_1\bar g_2$ carries
vanishing $P$ and $R$.  Thus $W_1$ can contain $ABB$, $BBH_1
H_2/M$, $\Psi^a\Psi^a$, etc.  On the other hand $g\bar g$ can
contain $A^*H_1H_2/M$, $A^*B^*B^*\Psi^a\Psi^a/M^3$, etc.  From
$W_1$ and $g_1\bar g_2$, we do not expect to generate a potential
containing $a_c$ and $a_h$ since these phase fields do not appear
in $\sum_{i}|\partial W_{\rm eff}/\partial z_i|^2$ and
$|W_{\rm eff}|^2$.
Thus $V_R(\theta_c,\theta_h)=$ (constant).  The constant is chosen
so that the potential is zero at the minimum of the potential.
Note, however, that $V_R$ can contain the scalar partners of
the phase fields.

To see the $a_c$ and $a_h$ independence of $V_R$, we observe that
$W_{\rm eff}$ must preserve $U(1)_P \times U(1)_R$, which is the case
for $U(1)_P$ if nonperturbative effects of gravitational interaction
are supposed to respect it.  Namely, we do not consider the argument
based on wormholes, otherwise the axion solution of the
strong CP problem is not attractive. Also, it is assumed that
the $U(1)_R$ symmetry in some string models is valid up to dim=10
operators.  This extra assumption is needed for the strong CP,
but not for a solution of the $\mu$ problem.
Suppose $z_i$ carries $P$ and $Q$ charges so
that it transforms under $U(1)_P$ and $U(1)_R$ as
\be
z_i\longrightarrow e^{i[\alpha_c P(z_i)+\alpha_h R(z_i)]}z_i
\ee
where $\alpha_c$ and $\alpha_h$ are the rotation angles.  The $U(1)_P$
and $U(1)_R$ symmetry of $W_{\rm eff}$ implies
\begin{eqnarray}
P: W_{\rm eff}\longrightarrow &W_{\rm eff} \\
R: W_{\rm eff}\longrightarrow & e^{i\alpha_h R} W_{\rm eff}
\end{eqnarray}
where $R=\sum_i R(z_i)=2$.  The nonlinear realization of this
transformation is represented by Goldstone fields $a_c$ and $a_h$.
$W_{\rm eff}$ does not depend on $a_c$ but depends on $a_h$, which
can be factored out as
\be
W_{\rm eff}\ =\ (W_{\rm eff})_{\rm radial}e^{2ia_h/v_h}
\ee
with the following transformation
\be
a_h \longrightarrow a_h+\alpha_h v_h
\ee
where $(W_{\rm eff})_{\rm radial}$ does not contain $a_h$.  $|W_{\rm eff}
|^2$ is independent of $a_c$ and $a_h$.  $\partial W_{\rm eff}/\partial
z_i$ carries $P=-P(z_i)$ and $R=2-R(z_i)$ and can be written
\be
\left({\partial W_{\rm eff}\over \partial z_i}\right)_{\rm radial}
\exp \left[i\left(-P(z_i){a_c\over v_c}+(2-R(z_i)){a_h\over v_h}\right)
\right]
\ee
where $(\partial W_{\rm eff}/\partial z_i)_{\rm radial}$ does not
depend on $a_c$ and $a_h$.  Therefore, $\sum_i |\partial W_{\rm eff}/
\partial z_i|^2$ does not depend on $a_c$ and $a_h$.

In general, we expect that $\nu_0$ and $\nu_h$
are different, and both $\theta_h$ and $\theta_c$ can be settled
to zero by the dynamical fields $a_h$ and $a_c$.  The mass matrix
$M^2$ of $a_c$ and $a_h$ satisfies
\be
Det\ M^2\ =\ (\nu_0-\nu_h)^2{\Lambda^4_{QCD}\Lambda^4_h\over v_c^2
v_h^2}
\ee
from which we expect an invisible axion, with
mass $\sim {\rm (coupling \ constant})\times
\Lambda_{QCD}^2/\Lambda_h$, because $\nu_0\ne \nu_h$
and $\Lambda_h$, $v_c$ and $v_h$ are of the same order.
The other boson gets a mass of order (coupling constants)$
\times \Lambda_h^2$.

The $\mu$ term given in Eq. (1) is supersymmetric.  In principle,
it can arise without supersymmetry breaking.  Thus it can arise
from a term $AH_1H_2$ in the superpotential with $\langle A\rangle
\ne 0$, and $\mu$ can be of order string scale.
In the above, we argued that a Peccei--Quinn symmetry spontaneously
broken at the intermediate mass scale $M_I$ excludes this
possibility.  Then there exist many possibilities for generating
the $\mu$ term of order of $M_{SUSY}$.

Recently, the $\mu$ term
has been calculated
in some string models\cite{anto} indicating the presence of terms as described
in the paper of Guidice and Masiero\cite{gm}. The contribution of
terms in the superpotential to the $\mu$ term will, however, be
dominant unless they are forbidden by a symmetry.

A general expression for the $\mu$ term can be written as

\be
\mu\ =\ A+m_{3/2}O(1)-F^{\bar n}\partial_{\bar n}
\xi (M^{\bar n})+\tilde \mu
\ee
where $F^n$ is the auxiliary component
of $n$-th modulus, and $M^{\bar n}$ is the anti-modulus\cite{anto}.
The $A$ term must be absent, presumably by
the Peccei--Quinn symmetry as mentioned above.  The
$m_{3/2}$ term arises from the Guidice--Masiero or the
Casas--Munoz form \cite{{gm},{munoz}}.  The rest arise from
 $D$ terms; thus involve anti-chiral fields.  $\xi$ arises
from moduli couplings and $\tilde \mu$ arises from the
Yukawa and gauge couplings.  In any case, these are at most
of order $m_{3/2}$, since to obtain a superpotential $W_\mu$
one must take $\int d^2\bar\theta$ on the $D$-term which
will pick up the $F$-term ($\sim M_I^2$) of anti-chiral
fields.

Notice, that the absence of $A$ in Eq. (28) is crucial
for the solution of the $\mu$ problem and has to be guaranteed by
a symmetry, as e.g.
the anomalous
$U(1)$ symmetry in the models  discussed above.

\noindent {\bf 5. Conclusion}

We have seen that there can be many sources for
the $\mu$ term in supergravity.  To understand its magnitude
in a natural scheme, however, one needs a symmetry. Both
the $R$ symmetry and the Peccei--Quinn symmetry are suggested for
a natural solution of the $\mu$ problem.   Without such a
symmetry principle, one does not
have a handle to remove specific (large) terms in the super-potential.
We have shown that string models with the anomalous
$U(1)_A$ with specific $P$ and $R$ charges are the best candidates
toward the solution of both the strong CP problem and the $\mu$
problem.

\acknowledgments
This work is supported in part by the Korea Science and Engineering
Foundation through Center for Theoretical Physics, Seoul
National University (JEK), KOSEF--DFG Collaboration Program (JEK, HPN),
 the Basic
Science Research Institute Program, Ministry of Education, 1994,
BSRI-94-2418 (JEK) and  European Union grants
SC1-CT91-0729 and SC1-CT92-0789 (HPN).

\end{document}